\documentclass[preprint]{elsarticle}
\usepackage{lineno,hyperref,pdflscape}
\modulolinenumbers[5]

\usepackage{multicol}
\usepackage{color}
\usepackage{xspace}
\usepackage{enumerate}
\usepackage{amsmath} 
\usepackage{ulem} 
\usepackage{placeins}
\usepackage{threeparttable}

\usepackage{tabularx}
\usepackage{mathtools} 
\usepackage{amsfonts}
\newcolumntype{M}{>{\centering\arraybackslash}m{1.85cm}}
\usepackage[export]{adjustbox}
\usepackage{float}
\usepackage{braket}
\usepackage{lipsum}
\usepackage{longtable}
\usepackage{lscape}
\graphicspath{{figure/}}
\newcommand\T{\rule{0pt}{3ex}}       
\newcommand\B{\rule[-1.5ex]{0pt}{0pt}} 

\usepackage{booktabs}
\usepackage{blindtext}
\makeatletter
\newcommand{\colorcaption}[2][]{%
	\begingroup%
	\renewcommand{\@caption@fignum@sep}{ (Color online). }%
	\caption[#1]{#2}%
	\endgroup%
}

\newcommand{\lambdabar}{{\mkern0.75mu\mathchar '26\mkern -9.75mu\lambda}}
\makeatother

\usepackage{amssymb}
\usepackage{graphicx}
\usepackage{rotating}
\usepackage{tikz}
\begin{document}

\begin{frontmatter}
\title{Shell-model study of $\log ft$ values for $^{139,140,141}$Ba $\rightarrow$ $^{139,140,141}$La  transitions}

\author{Shweta Sharma}
\author{Praveen C. Srivastava\footnote{Corresponding author: praveen.srivastava@ph.iitr.ac.in}}
\address{Department of Physics, Indian Institute of Technology Roorkee, Roorkee 247667, India}

\date{\hfill \today}
\begin{abstract}

In the present work, beta-decay properties such as $\log ft$ values and half-lives have been systematically studied corresponding to Ba isotopes using large-scale shell-model calculations. An extensive comparison of beta decay results corresponding to $^{141}$Ba$\rightarrow$ $^{141}$La using shell-model calculations is made with the recently available experimental data. In addition, we have also calculated the nuclear and beta decay properties corresponding to $^{139}$Ba$\rightarrow$ $^{139}$La and $^{140}$Ba$\rightarrow$ $^{140}$La transitions.  The model-space considered here is $Z=50-82$ and $N=82-126$ with $^{132}$Sn core, and the interaction employed here is jj56pnb interaction. The beta decay results using shell-model calculations for all the mentioned isotopes are compared with the available experimental data. This is the first theoretical interpretation corresponding to recent experimental data.

\end{abstract}

\end{frontmatter}


\section{Introduction}
The beta decay study and its theoretical description are important for better estimation of nuclear matrix elements \cite{suhonen1998,suhonen2005}.   In addition, the beta decay study provides a deeper insight into the nuclear structure and nuclear interactions involved in the weak interaction processes. 
Thus, one needs a theoretical model that gives an accurate and thorough description of the beta decay theory. There are various theoretical models that have been utilized for this purpose, and in the present work, we are using the interacting shell-model \cite{dean2004,caurier2005}.  The nuclear shell-model is very beneficial in the calculation of several important nuclear properties and matrix elements involved in beta decay processes, which gives information about the nuclear structure and corresponding wave functions. The study of beta decay processes is very useful in understanding astrophysical processes.
In Ref. \cite{Sieja}, shell-model results for neutron capture rates for $fp$ shell nuclei are reported. The nuclear properties for nuclear astrophysical studies have recently been reported in Ref. \cite{Goriely}. 


The shell-model calculations usually overestimate the beta-decay strength functions \cite{zhi2013,martinez1996}. Thus, these strength functions are calculated using quenching factors in the weak coupling constants \cite{zhi2013,martinez1996,brown1985,suhonen2017,joel12017,joel22017,kumar2016jpg,kumar2016,kumar2023}. Also, the mesonic enhancement factor \cite{towner1992} is needed in the $0^-$ transitions of the beta decay using shell-model calculations because of impulse approximation. This mesonic enhancement factor is evaluated for different regions in the nuclear chart \cite{warburton11991,warburton21991,warburton1988,warburton1994,kostensalo2018}. 
Further, the competition between the first forbidden and allowed transitions in the heavier mass region from recent experimental data from CERN is reported earlier in Refs. \cite{carroll2020,brunet2021}.
Keeping all these factors in mind, earlier we have studied beta-decay properties of nuclei in the $^{208}$Pb region using large-scale shell-model calculations \cite{sharma2022,srivastava2022,anil2021,sharma2023arxive}. Since the nuclei in the $^{132}$Sn region show similarity in nuclear properties with the nuclei in the $^{208}$Pb region \cite{coraggio2009,bhoy2020}. Thus, we have also extended our study towards the $^{132}$Sn region and calculated beta decay properties of $^{135,137}$Te nuclei using shell-model calculations \cite{sharma2023}. Our group has also studied beta decay properties for sd shell nuclei \cite{anil2020,anil2020ptep,priyanka2021}  using shell-model calculations.
 
Recently, several experiments have been carried out to study beta decay properties for allowed and forbidden transitions in the different mass regions \cite{suhonen2017}. In the Sn region, experiments have been carried out to
 investigate the excited levels of La isotopes. The results corresponding to beta decay of $^{139}$Ba to $^{139}$La are reported in Refs. \cite{berzins1969,faller1986,zamboni2001,hill1968,gehrke1980,danu2012}. The decay scheme of $^{139}$Ba has been studied by using NaI(Tl)-Ge(Li) detectors by  Berzins  \textit{et al} in Ref. \cite{berzins1969}. Later, this decay scheme along with the decay scheme of $^{141}$Ba was studied  by Faller \textit{et al} \cite{faller1986}. They observed that there is no energy level between 166 and 1209 keV in $^{139}$La. Later, Zamboni \textit{et al} \cite{zamboni2001} proposed two new levels at 1524.6 and 1900.3 keV excitation energies corresponding to beta decay $^{139}$Ba. Recently, Danu  \textit{et al} \cite{danu2012} investigated the decay scheme of $^{139}$Ba using singles and $\gamma-\gamma$ coincidence spectroscopy techniques with four Compton suppressed Clover HPGe detectors at BARC, Mumbai. In this work, various measurements have been taken like energy spectra, $\log ft$ values and half-life. Further, beta decay of $^{140}$Ba to $^{140}$La has been investigated by Meyer \textit{et al} \cite{meyer1990} by populating the levels of $^{140}$La using $\gamma$-ray spectroscopy.
 
Greenwood \textit{et al} \cite{greenwood1997} investigated the beta decay corresponding to $^{141}$Ba by measuring beta feeding intensities. Due to the lack of errors in their beta feeding intensities results, Rufino \textit{et al} \cite{rufino2022} performed a high-sensitivity, high-resolution $\gamma$-ray measurements to investigate the $^{141}$Ba to $^{141}$La beta decay. They carried out these beta decay measurements at Argonne National Laboratory (ANL) using the Gammasphere array and measured the beta decay strength functions which were compared with the $\log ft$ results earlier measured by Faller \textit{et al} \cite{faller1986}. However, no theoretical interpretation has been performed.

 Thus, motivated by this new experimental data we have performed a systematic shell-model study for the beta decay properties of $^{141}$Ba to $^{141}$La transitions. Along with this we have also studied $^{139,140}$Ba to $^{139,140}$La transitions. 

This work is organized as follows:
Section \ref{formalism} is concerned with the formalism behind allowed and forbidden beta decay and Section \ref{results} shows the results and discussion of the present work which includes the calculation of nuclear structure and beta decay properties corresponding to the aforementioned nuclei. This discussion is followed by the conclusion in section \ref{conclusion} at the end. 

\section{Formalism}\label{formalism}
\subsection{Shell-model Hamiltonian}
The shell-model Hamiltonian \cite{suhonen2007} in terms of single-particle energies and two-body interaction between the nucleons can be defined as
\begin{equation}
H = T + V = \sum_{\alpha}{\epsilon}_{\alpha} c^{\dagger}_{\alpha} c_{\alpha} + \frac{1}{4} \sum_{\alpha\beta \gamma \delta}v_{\alpha \beta \gamma \delta} c^{\dagger}_{\alpha} c^{\dagger}_{\beta} c_{\delta} c_{\gamma},
\end{equation}
where the term $\alpha = \{n,l,j,t\}$ denotes the single-particle orbitals. The quantity $\epsilon_{\alpha}$ is the single particle energy of the state $\alpha$. The terms $c^{\dagger}_{\alpha}$ and $c_{\alpha}$ are creation and annihilation operators, respectively. While, $v_{\alpha \beta \gamma \delta} = \langle\alpha \beta | V | \gamma \delta\rangle $ are the antisymmetrized two-body matrix elements (TBMEs).
\subsection{Beta decay formalism}
The streamlined formalism for allowed and forbidden beta decay is given here, which is based on the concept of impulse approximation. According to impulse approximation, the decaying nucleon only feels the weak interaction at the instant of decay and does not interact with the remaining nucleons. It feels the strong interaction with the remaining nucleons only before and after the decay. The detailed formalism can be found in Refs. \cite{suhonen2007,behrens1982,schopper1996}.
 \begin{figure}
 \begin{center}

 \includegraphics[scale=1.7]{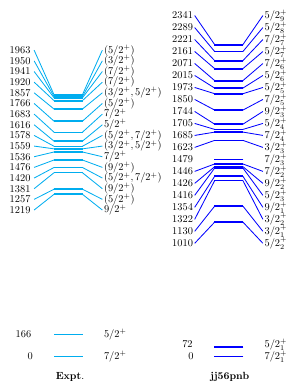}
	\label{spectra1}
			\caption{Comparison between experimental \cite{joshi2016} and shell-model energy spectra of $^{139}$La with corresponding energies are in keV.\label{spectra1}}
        
 \end{center}
\end{figure}

The partial half-life for a transition in beta decay is given by
\begin{eqnarray}\label{hf1}
	t_{1/2}=\frac{\kappa}{\tilde{C}},
	\end{eqnarray}
where $\kappa=6289$ s is the constant value \cite{patrignami2016} and $\tilde{C}$ is the unitless integrated shape factor, given by
 \begin{eqnarray} \label{tc}
	\tilde{C}=\int_1^{w_0}C(w_e)pw_e(w_0-w_e)^2F_0(Z,w_e)dw_e,
	\end{eqnarray}
 where $p$ and $w_e$ are the electron momentum and energy, respectively. The quantity $w_0$ is the end-point energy, i.e., the highest amount of energy acquired by the emitting electron. The term $C(w_e)$ is the shape factor and $F_0(Z,w_e)$ is the Fermi function for the daughter nucleus with proton number $Z$. The shape factor $C(w_e)$ can be written as
	\begin{eqnarray} \label{eqn4}
	\begin{split}
	C(w_e)  = \sum_{k_e,k_\nu,K}\lambda_{k_e} \Big[M_K(k_e,k_\nu)^2+m_K(k_e,k_\nu)^2 \\
	-\frac{2\gamma_{k_e}}{k_ew_e}M_K(k_e,k_\nu)m_K(k_e,k_\nu)\Big].
	\end{split}
	\end{eqnarray} 
 \begin{figure}
\begin{center}
		\includegraphics[scale=1.7]{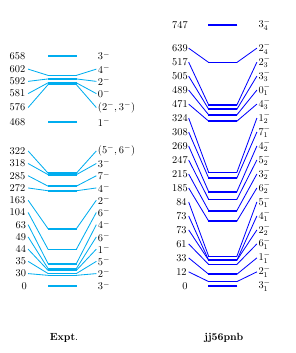}
	\label{spectra2}
			\caption{Comparison between experimental \cite{nica2018} and shell-model energy spectra of $^{140}$La with corresponding energies are in keV.\label{spectra2}}
   \end{center}
\end{figure}
	Here, the lowest feasible value of $K$, or forbiddenness order, is chosen since the terms with the least angular momentum transfer account for the majority of contributions.  As a result, $K=K_{min},K_{min}+1$ where $K_{min}=|J_i-J_f|$ with $J_i$ and $J_f$ represents the initial and final angular momentum, respectively. The terms $k_e$ and $k_\nu$ stand for the positive integers that result from the partial wave expansion of the electron and neutrino wave functions, respectively.  Given that contributions from higher order terms in the leptonic wave function expansion are minimal, there are two ways to get the total $k_e+k_{\nu}$ to be as small as possible: $k_e+k_{\nu}=K+1$ and $k_e+k_{\nu}=K+2$. The quantities $M_K(k_e,k_\nu)$ and $m_K(k_e,k_\nu)$ are the complicated combinations of the different form factors incorporating nuclear structure information and the leptonic phase space factors. Additional details on these quantities can be found in \cite{haaranen2017,mustonen2006}. The impulse approximation can be used to relate the nuclear form factors to the corresponding nuclear matrix elements (NMEs), given by
 
 \begin{figure}
\begin{center}
		\includegraphics[scale=1.7]{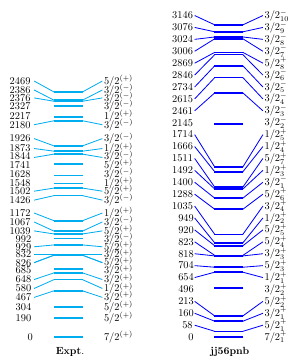}
	\label{spectra3}
			\caption{Comparison between experimental \cite{rufino2022} and shell-model energy spectra of $^{141}$La with corresponding energies are in keV.}
   \label{spectra3}
   \end{center}
\end{figure}

\begin{table}
	 \centering
	\caption{The rms deviations between experimental and shell-model energy levels corresponding to $^{139,140,141}$La isotopes. }\label{rms} 
	\begin{tabular}{ccccccc}
 \hline\hline
		Isotope& \multicolumn{2}{c}{~~~rms deviation (MeV)}\T\B \\ \hline
	    $^{139}$La&~~~~~~~~~~0.164 \\ \T\B
		$^{140}$La&~~~~~~~~~~0.0733 \\ \T\B
		$^{141}$La&~~~~~~~~~~0.558\\\hline\hline
		
	\end{tabular}
\end{table}

\begin{eqnarray} 
 \begin{split}
 ^{V/A}\mathcal{M}_{KLS}^{(N)}(pn)(k_e,m,n,\rho)& \\ 
 =\frac{\sqrt{4\pi}}{\widehat{J}_i}
 \sum_{pn} \, ^{V/A}m_{KLS}^{(N)}(pn)(&k_e,m,n,\rho)(\Psi_f\parallel [c_p^{\dagger}\tilde{c}_n]_K\parallel \Psi_i),
 \end{split}
 \end{eqnarray} 

where $m$ is the total power of $(m_eR/\hbar)$, $(W_eR/\hbar)$, and $\alpha Z$, $n$ is the total power of $(W_eR/\hbar)$ and $\alpha Z$, and $\rho$ is the power of $\alpha Z$ and ${\widehat{J}_i}=\sqrt{2J_i+1}$ and the summation is given over protons and neutrons single-particle states. The model-independent term ${^{V/A}m_{KLS}^{(N)}}(pn)(k_e,m,n,\rho)$ is defined as the single particle matrix elements (SPMEs), whereas the model dependent term $(\Psi_f|| [c_p^{\dagger}\tilde{c}_n]_K || \Psi_i)$ is the one body transition density (OBTD) with $(\Psi_i)$ and $(\Psi_f)$ being the initial and final nuclear states, respectively. The OBTDs along with other nuclear structure properties are obtained from the shell-model diagonalization codes NuShellX \cite{nushellx} and KSHELL \cite{kshell}. Further, the auxiliary quantities $\gamma_{k_e}$ and $\lambda_{k_e}$ (Coulomb function) in Eqn. (\ref{eqn4}) are given as 
	\begin{align*}
		\gamma_{k_e} & =\sqrt{k_e^2-(\alpha{Z})^2}&\mbox{and} \,\,\,\, \lambda_{k_e} & ={F_{k_e-1}(Z,w_e)}/{F_0(Z,w_e)},
	\end{align*} 
 where the quantity $\alpha=1/137$ is the fine structure constant. $F_{k_e-1}(Z,w_e)$ is the generalized Fermi function which is given as
 \begin{eqnarray}
	F_{k_e-1}(Z,w_e) &=4^{k_e-1}(2k_e)(k_e+\gamma_{k_e})[(2k_e-1)!!]^2e^{\pi{y}} \nonumber \\
	& \times\left(\frac{2p_eR}{\hbar}\right)^{2(\gamma_{k_e}-k_e)}\left(\frac{|\Gamma(\gamma_{k_e}+iy)|}{\Gamma(1+2\gamma_{k_e})}\right)^2,
	\end{eqnarray}
 where $y=(\alpha{Zw_e}/p_ec)$. 
 In the case of allowed beta decay, the average shape factor becomes independent of electron energy and is given by $C(w_e)=B(\text{F})+B(\text{GT})$ where $B(F)$ and $B(GT)$ are Fermi and Gamow-Teller reduced transition probabilities \cite{brown1985}. Further, in $0^-$ transitions the axial-vector matrix element gets improved over the impulse approximation with the aid of mesonic enhancement factor, i.e., $\epsilon_\textrm{MEC}$ \cite{towner1992}. The value of the mesonic enhancement factor is taken to be 2.0 in our work from Ref. \cite{kostensalo2018}.

	\begin{table*}
	\centering
	\caption{ Comparison between experimental \cite{nndc,iaea} and shell-model calculated electromagnetic observables corresponding to $^{139,140,141}$La. In the shell-model calculations, we have taken $e_p=1.5e, e_n=0.5e$ and $g_s^{eff}=g_s^{free}$.}\label{table2} 
		\begin{tabular}{ccccccccccc}
		\hline\hline
			Nucleus&\multicolumn{2}{c}{~~~~~$\mu(\mu_N)$} & \multicolumn{2}{c}{~~~~~$Q(eb)$} \T\B \\
			\cline{2-3}
			\cline{4-5}
			&	Expt. & SM&  Expt. & SM      \T\B\\\hline\hline\T\B
			$^{139}$La($7/2^+_1$)	&2.7830455(9) & 1.760&  +0.200(6) & +0.127\T\B\\\hline\T\B
			$^{140}$La($3^-_1$)	&0.730(15) & 0.130&  +0.094(10) & +0.152\T\B\\\hline\T\B
   $^{141}$La($7/2^+_1$)	&NA & 1.726&  NA & +0.030\T\B\\\hline\T\B
			&\multicolumn{2}{c}{~~~~~$ B(M1)(\mu_N^2)$} &  \multicolumn{2}{c}{~~~~~$B(E2)(e^2fm^4)$}\T\B \\
			\cline{2-3}
			\cline{4-5}
			&	Expt. & SM&  Expt. & SM      \T\B\\\hline\hline\T\B
			$^{139}$La($5/2^+_1 \rightarrow 7/2^+_1$)	 & 0.00460(7) & 0.0003&$<$85.54 & 1.62\T\B\\\hline\T\B
   $^{140}$La($2^-_1 \rightarrow 3^-_1$)	& 0.913$^{+197}_{-143}$ & 0.813 & $<$1434 & 349\T\B\\\hline\T\B
			$^{141}$La($5/2^+_1 \rightarrow 7/2^+_1$)	& 0.00381$^{+34}_{-18}$ & 0.005& NA & 137\\\hline\hline
		\end{tabular}
\end{table*}

 \begin{figure*}
 \centering
		\includegraphics[scale=0.96]{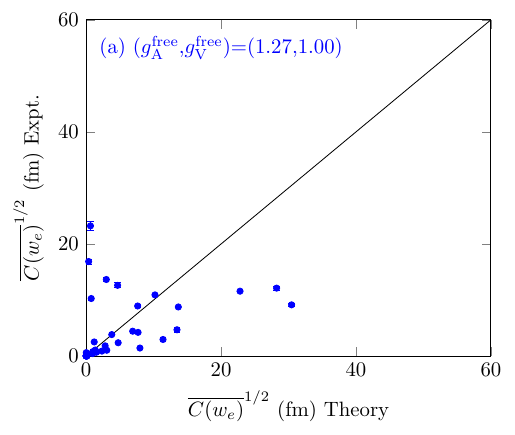}
		\includegraphics[scale=0.96]{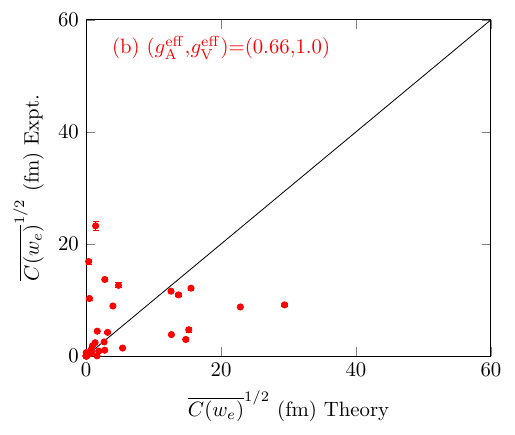}
			\caption{Comparison of experimental \cite{rufino2022,nndc} and calculated average shape factor calculated via (a) bare values of weak coupling constants (b) effective values of weak coupling constants.\label{shape_plot}}
		\end{figure*}

 The half-life for a specific transition can vary in a wide range. Thus, it is generally given in terms of $\log ft$ values.
 \begin{equation}\label{ft}
	\log ft \equiv \log(f_{0}t_{1/2}).
\end{equation}

In this expression, $t_{1/2}$ can be obtained from Eq. (\ref{hf1}) and the phase space factor $f_0$ can be expressed as
	\begin{equation}
	f_{0}= \int_1^{w_0}pw_e(w_0-w_e)^2F_0(Z,w_e)dw_e.
	\end{equation}

 Further, in the case of forbidden unique transitions, the phase space factor is given by $f_{Ku}$ which is expressed as
 
	\begin{align}
		&f_{Ku}= \Big(\frac{3}{4}\Big)^K\frac{(2K)!!}{(2K+1)!!} \nonumber \\ &\times\int_1^{w_0}C_{Ku}(w_e)p w_e(w_0-w_e)^2F_0(Z_f,w_e)dw_e,
	\end{align}
	where $C_{Ku}(w_e)$ is the shape function for $K^{\text{th}}$ forbidden unique transition which is expressed as
	\begin{equation}
	C_{Ku}(w_e)=\sum_{k_e+k_{\nu}=K+2}\frac{\lambda_{k_e}p^{2(k_e-1)}(w_0-w_e)^{2(k_{\nu}-1)}}{(2k_e-1)!(2k_{\nu}-1)!} .
	\end{equation}
	
and $f_{1u}=12 f_{K=1,u}$ for the first forbidden unique transitions.

	Further, the experimental Gamow-Teller strength and forbidden transition probabilities are usually smaller in value as compared to the values calculated with the help of the shell-model because of various factors such as truncated model-space, nuclear medium effects, and impulse approximation used in the beta decay formalism \cite{suhonen2017}. Thus, there is a need to quench these values. This quenching is performed via the renormalization of the vector and axial vector coupling constant and is calculated by comparing theoretical and experimental average shape factors. These average shape factor values are given as
 \begin{equation}
	\overline{C(w_{e})} (\text{fm}^{2K}) =\frac{6289 
        \lambdabar_{\text{Ce}}^{2K}}{ft},
\end{equation}
where $ \lambdabar_{\text{Ce}}=386.159$ fm stands for the reduced Compton wavelength of the electrons and $ft$ is obtained from the Eq. (\ref{ft}).

\begin{landscape}
\begin{table*}
	 \centering
	\caption{ Comparison between theoretical and experimental \cite{joshi2016} $\log ft$ values for $^{139}$Ba$(7/2^-)$ to the different excited states in $^{139}$La($J_f$) transitions using effective value of axial-vector coupling constant i.e. $(g_A^{eff}=0.66\pm 0.03)$ and $\epsilon_{MEC}=2.0$.\label{table1}} 
	\begin{tabular}{lccccccc}
 \hline\hline
	\multicolumn{2}{c}{~~~~Final state} & Decay mode & Energy (keV)&\multicolumn{2}{c}{~~~~~$\log ft$} & \multicolumn{2}{c}{~~~~~$ {[\overline{C(w_{e})}]}^{1/2}$} \T\B \\
		\cline{1-2}
		\cline{5-6}
		\cline{7-8}
		
		$J_f^{\pi}$ (Expt.) & $J_f^{\pi}$ (SM) & & & Expt. & SM & Expt. & SM     \T\B\\\hline\hline\T\B
		$7/2^+$&$7/2^+_1$&1st FNU & 0.0&6.845(3)	&6.774 & 11.576(40)& 12.562 \\ \T\B
		$5/2^+$&$5/2^+_1$ & 1st FNU & 165.859(7)&7.087(5)	&6.254 & 8.761(50)& 22.859\\ \T\B
		$9/2^+$&$9/2^+_1$& 1st FNU&1219.047(10) & 9.68(1) & 9.332 & 0.443(5) &0.661  \\ \T\B
		$(5/2)^+$&$5/2^+_2$&1st FNU	& 1256.797(11)&10.24(1) & 8.557 & 0.232(3) & 1.613 \\ \T\B
		$(9/2^+)$ & $9/2^+_2$ & 1st FNU	& 1381.409(14)&9.65(1)	&8.689 &0.458(5)  & 1.385 \\ \T\B
		$(5/2^+,7/2^+)$ &$5/2^+_3$ & 1st FNU& 1420.491(8)&7.64(1) & 8.708 & 4.635(53) & 1.355  \\ \T\B
		$(5/2^+,7/2^+)$ &$7/2^+_2$ & 1st FNU&1420.491(8)& 7.64(1)	& 7.805 & 4.635(53) & 3.833 \\ \T\B
		$(9/2^+)$& $9/2^+_3$&1st FNU &1476.489(8)	& 8.78(2)	& 10.763 &1.248(29) &  0.127\\ \T\B
		$7/2^+$&   $7/2^+_3$&1st FNU 	&1536.388(8)& 9.19(2) & 9.208&0.778(18) &  0.762\\ \T\B
		$(3/2^+,5/2^+)$ &$3/2^+_1$ & 1st FU	& 1558.72(3)&10.21(2)	& 9.371 & 0.240(6)& 0.632  \\ \T\B
		$(3/2^+,5/2^+)$ &$5/2^+_4$ & 1st FNU &1558.72(3)& 10.21(2)	& 9.292 & 0.240(6) & 0.692\\ \T\B
		$(5/2^+,7/2^+)$ &$5/2^+_5$ &1st FNU	& 1578.156(14)&10.00(1)	& 9.317 & 0.306(4) & 0.672  \\ \T\B
		$(5/2^+,7/2^+)$ &$7/2^+_4$ & 1st FNU & 1578.156(14)& 10.00(1)	& 8.322 & 0.306(4) & 2.114\\ \T\B
		$7/2^+$&  $7/2^+_5$	& 1st FNU&1683.145(10)& 9.08(1) & 8.431 & 0.883(10) & 1.864 \\ \T\B
		$(5/2)^+$&  $5/2^+_7$	& 1st FNU&1766.432(25)& 9.59(2) & 9.660 & 0.491(11)  & 0.453\\ \T\B
		$(3/2^+,5/2^+)$ &$3/2^+_2$ & 1st FU	& 1856.62(4)&10.1(3)	& 10.485 &0.273(94) & 0.175  \\ \T\B
		$(3/2^+,5/2^+)$ &$5/2^+_8$ & 1st FNU &1856.62(4)& 10.1(3)	& 10.801 & 0.273(94) & 0.122\\ \T\B
		$(7/2^+)$&$7/2^+_6$& 1st FNU &1920.43(4)& 9.43(2)	& 8.894 & 0.590(14) & 1.094 \\ \T\B
		$(5/2^+)$&  $5/2^+_9$	& 1st FNU&1962.84(11)& 9.88(6) & 10.187 & 0.352(24) & 0.247\\\hline\hline
		\end{tabular}
\end{table*}

\begin{table*}
	\centering
	\caption{ Comparison between theoretical and experimental \cite{nica2018} $\log ft$ values for $^{140}$Ba$(0^+)$ to the different excited states in $^{140}$La($J_f$) 
 transitions using effective value of axial-vector coupling constant i.e. $(g_A^{eff}=0.66\pm 0.03)$ and $\epsilon_{MEC}=2.0$. Here, '$u$' denotes forbidden unique decay.\label{table2}} 
	\begin{tabular}{lccccccc}
 \hline\hline
		\multicolumn{2}{c}{~~~~Final state} & Decay mode & Energy (keV)&\multicolumn{2}{c}{~~~~~$\log ft$} & \multicolumn{2}{c}{~~~~~$ {[\overline{C(w_{e})}]}^{1/2}$} \T\B \\
		\cline{1-2}
		\cline{5-6}
		\cline{7-8}
		
		$J_f^{\pi}$ (Expt.) & $J_f^{\pi}$ (SM) & & & Expt. & SM & Expt. & SM     \T\B\\\hline\hline
			$2^-$&$2^-_1$&1st FU & 29.9677(8)&8.91$^{1u}$(24)	&9.272$^{1u}$ &1.074(297) & 0.708 \\ \T\B
		$1^-$&$1^-_1$ & 1st FNU & 43.8132(18)&8.03(11)	&6.634 &2.958(375) & 14.759 \\ \T\B
		$2^-$&$2^-_2$& 1st FU&162.6585(8) & 9.30$^{1u}$(3) & 9.166$^{1u}$ & 0.686(24) &0.800 \\ \T\B
		$1^-$&$1^-_2$&1st FNU	& 467.5350(17)&7.805(23) & 6.769 & 3.833(102) & 12.634 \\ \T\B
		$0^-$ & $0^-_1$ & 1st FNU	& 581.073(9)&7.07(3)	& 7.775 & 8.934(309) & 3.968  \\\hline\hline
	\end{tabular}
\end{table*}

  \begin{table*}
	 \centering
	\caption{ Comparison between theoretical and experimental \cite{rufino2022} $\log ft$ values for $^{141}$Ba$(3/2^-)$ to the different excited states in $^{141}$La($J_f$) transitions using effective value of axial-vector coupling constant i.e. $(g_A^{eff}=0.66\pm 0.03)$ and $\epsilon_{MEC}=2.0$. Here, '$u$' denotes forbidden unique decay.\label{table3}} 
	\begin{tabular}{lccccccc}
 \hline\hline
		\multicolumn{2}{c}{~~~~Final state} & Decay mode & Energy (keV)&\multicolumn{2}{c}{~~~~~$\log ft$} & \multicolumn{2}{c}{~~~~~$ {[\overline{C(w_{e})}]}^{1/2}$} \T\B \\
		\cline{1-2}
		\cline{5-6}
		\cline{7-8}
		
		$J_f^{\pi}$ (Expt.) & $J_f^{\pi}$ (SM) & & & Expt. & SM & Expt. & SM     \T\B\\\hline\hline
		$7/2^{(+)}$&$7/2^+_1$&1st FU & 0.0&9.4$^{1u}$(5)	&12.974$^{1u}$ &0.611(352) & 0.010 \\ \T\B
		$5/2^{(+)}$&$5/2^+_1$ & 1st FNU & 190.40(20)&7.63(8)	&6.608&4.689(432) & 15.207 \\ \T\B
		$3/2^{(+)}$&$3/2^+_1$& 1st FNU&467.37(8) & 7.05(3) & 6.035 & 9.142(316) &29.414  \\ \T\B
		$1/2^{(+)}$&$1/2^+_1$&1st FNU	& 580.17(10)&8.45(5) & 9.033  & 1.824(105) & 0.932 \\ \T\B
		
		$3/2^{(+)}$ & $3/2^+_2$ & 1st FNU	& 647.94(8)&6.805(24)	& 6.589 & 12.122(335) & 15.544 \\ \T\B
		$3/2^{(+)}$ &$3/2^+_3$ & 1st FNU& 685.40(10)&8.66(3) & 7.507 & 1.432(49) & 5.402  \\ \T\B
		$5/2^{(+)}$ &$5/2^+_3$ & 1st FNU&826.42(10)	& 8.17(3)	& 8.121 & 2.518(87) & 2.664 \\ \T\B
		$3/2^{(+)}$& $3/2^+_4$&1st FNU &831.66(8)	& 7.720(24)	& 7.969 &4.227(117) &  3.174\\ \T\B
		$5/2^{(+)}$&   $5/2^+_4$&1st FNU 	&929.44(9)& 6.896(25) & 6.699&10.916(314) &  13.695 \\ \T\B
		$3/2^{(-)}$ & $3/2^-_1$ & Allowed	& 991.97(10)&9.7(3)	& 6.277 & 0.0011(4) & 0.058  \\ \T\B
		$5/2^{(+)}$&$5/2^+_5$& 1st FNU &1039.48(9)& 8.94(6)	&8.091 & 1.038(72) & 2.758\\ \T\B
		$3/2^{(-)}$&  $3/2^-_2$	& Allowed&1066.57(10)& 7.317(25) & 6.373 & 0.017(1) & 0.052 \\ \T\B
		$1/2^{(+)}$&$1/2^+_2$& 1st FNU &1171.99(9)& 7.678(24)	& 8.540 & 4.437(123)& 1.645 \\ \T\B
		$3/2^{(-)}$&  $3/2^-_3$	& Allowed&1426.36(10)& 7.910(25) & 8.073 & 0.0088(2) & 0.007 \\ \T\B
		$5/2^{(+)}$&$5/2^+_6$& 1st FNU &1501.56(9)& 6.701(23)	& 8.091 & 13.663(362) & 2.758\\ \T\B
		$1/2^{(+)}$&  $1/2^+_3$	& 1st FNU&1547.69(17)& 8.22(3) & 8.726 & 2.377(82) & 1.328\\ \T\B
		$3/2^{(-)}$& $3/2^-_4$& Allowed	& 1628.16(10)&6.899(24)	& 9.745 & 0.028(1) &  0.001 \\ \T\B
		$5/2^{(+)}$&  $5/2^+_7$ & 1st FNU	&1740.69(10)& 6.949(24)	& 9.602 &10.270(284)  & 0.484  \\ \T\B
		$3/2^{(-)}$&  $3/2^-_5$ &Allowed 	& 1844.30(11)&6.749(24)	&8.606& 0.035(1) & 0.004 \\ \T\B
		$1/2^{(+)}$&  $1/2^+_4$ & 1st FNU	&1872.60(9)& 6.518(24)	& 9.845 &16.868(466)  & 0.366  \\ \hline\hline
\end{tabular}
\end{table*}

\addtocounter{table}{-1}
	\begin{table*}
		\caption{ Continued.}
 \centering
		\begin{tabular}{lccccccc}
 \hline\hline
		\multicolumn{2}{c}{~~~~Final state} & Decay mode & Energy (keV)&\multicolumn{2}{c}{~~~~~$\log ft$} & \multicolumn{2}{c}{~~~~~$ {[\overline{C(w_{e})}]}^{1/2}$} \T\B \\
		\cline{1-2}
		\cline{5-6}
		\cline{7-8}
  		$J_f^{\pi}$ (Expt.) & $J_f^{\pi}$ (SM) & & & Expt. & SM & Expt. & SM     \T\B\\\hline\hline
		$3/2^{(-)}$&  $3/2^-_6$ &Allowed 	& 1926.01(10)&6.724(25)	&7.000& 0.034(1) & 0.025\\ \T\B
		$3/2^{(-)}$&  $3/2^-_7$ & Allowed	&2180.38(12)& 6.964(25)	& 10.241 &0.026(1)  & 0.001  \\ \T\B
		$1/2^{(+)}$&  $1/2^+_5$ &1st FNU 	& 2216.56(11)&6.77(3)	&7.610& 12.620(436) & 4.798 \\ \T\B
		$3/2^{(-)}$& $3/2^-_8$& Allowed	& 2327.22(14)&7.06(3)	& 8.927  & 0.023(1) &  0.003 \\ \T\B
		$3/2^{(-)}$&  $3/2^-_9$ & Allowed	&2375.85(12)& 6.46(3)	& 8.570 &0.047(2)  & 0.004   \\ \T\B
		$3/2^{(-)}$&  $3/2^-_{10}$ &Allowed 	& 2385.68(11)&6.87(3)	&7.379& 0.029(1) & 1.621 \\ \T\B
		$5/2^{(+)}$&  $5/2^+_8$ & 1st FNU	&2468.74(9)& 6.24(3)	& 8.677 &23.230(802)  & 1.405\\\hline\hline
	\end{tabular}
 \end{table*}
\end{landscape} 			
  			
 \section{Results and discussion}\label{results}

In this section, the nuclear structure and beta decay properties corresponding to $^{139,140,141}$Ba $\rightarrow$ $^{139,140,141}$La transitions are reported using large-scale shell-model calculations. For these shell-model calculations, a particular model-space is chosen above $^{132}$Sn core. The model-space considered here includes five orbitals for protons, i.e., $0g_{7/2}$, $1d_{5/2}$, $1d_{3/2}$, $2s_{1/2}$, $0h_{11/2}$ ($50\leq Z \leq 82$) and six orbitals for neutrons, i.e.,  $0h_{9/2}$, $1f_{7/2}$, $1f_{5/2}$, $2p_{3/2}$, $2p_{1/2}$, $0i_{13/2}$ ($82 \leq N \leq 126$).  The effective interaction used for the above model space is the jj56pnb interaction \cite{kostensalo2018,sharma2023,brown}, which includes the Coulomb interaction part in the proton-proton interaction part.
Now using this effective interaction, first we have calculated the nuclear structure properties such as energy spectra and electromagnetic observables. Further, using the obtained wave functions we have performed calculations for the beta decay properties such as $\log ft$, average shape factor values, and half-lives. The average shape factors have been calculated using both bare and effective values of weak coupling constants. The comparison of experimental and shell-model average shape factor values is shown in this work for both bare and effective values of weak coupling constants. The effective value of weak axial-vector coupling constant ($g_A^{eff}=0.66\pm 0.03$) is taken from Ref. \cite{kostensalo2018} to calculate $\log ft$ values and half-lives in the present work. 
 These beta decay properties, such as the $\log ft$, average shape factor values, and half-lives are reported in this work and compared with the corresponding experimental data. The deviation in the shell-model results with respect to the experimental data is also reported in this section.

 Figs. \ref{spectra1}, \ref{spectra2} and \ref{spectra3} shows the energy spectra for ground and excited states of $^{139,140,141}$La isotopes, respectively using jj56pnb interaction and comparison with the experimental spectra for the same is also shown in these figures. In Fig. \ref{spectra1}, shell-model and experimental energy spectra for ground and excited states of $^{139}$La are shown. It can be seen from the figure that the jj56pnb interaction correctly reproduces the ground state (g.s.), i.e., $7/2_1^+$, and the first excited state as $5/2_1^+$. Further, there is a large gap between 72  and 1010 keV energy levels similar to the gap between 166 and 1219 keV experimentally.  In Fig. \ref{spectra2}, the experimental and shell-model energy spectra corresponding to the ground and excited states of $^{140}$La is shown. In this spectra, the ground state, i.e., $3_1^-$ and other excited states are correctly reproduced by jj56pnb interaction. For instance, the first excited state is $2_1^-$ experimentally, and the same is reproduced by our shell-model calculation. Similarly, Fig. \ref{spectra3} shows the experimental and shell-model energy spectra corresponding to the ground and excited states of $^{141}$La isotope. It also correctly reproduced the ground state and reasonably other excited states.  Furthermore, the root mean square (rms) deviations between the  experimental and the shell-model energy levels corresponding to $^{139,140,141}$La isotopes are computed in order to quantify the predictive power of jj56pnb interaction in the shell-model calculations. The root mean square deviation can be defined as
 \begin{equation}
	{\rm rmsd}=\Big[\frac{1}{n}\sum_{i=1}^n( E_i({\rm expt})-E_i({\rm SM}))^2\Big]^{1/2},
\end{equation}
where $E_i({\rm expt})$ is the experimental energy and $E_i({\rm SM})$ is the shell-model energy of the corresponding state. Here, $n$ is the total number of states considered for the calculation of rms deviation.
These rms deviations are computed for non-degenerate states only, and they are reported in Table \ref{rms}. 
While calculating the rms deviation for $^{139}$La, the states up to 1476 keV experimental energy are taken into account and the rms deviation comes out to be 0.164 MeV. Similarly, in the case of $^{140}$La, the $2^-$ and $3^-$ states at experimental energies 592 and 658 keV are not considered for rms deviation calculation because these states are above unconfirmed states $(2^-,3^-)$ at energy 576 keV. The rms deviation for $^{140}$La is 0.0733 MeV, and it is maximum for $^{141}$La, i.e, 0.558 MeV.

In addition to the energy spectra, the electromagnetic observables such as magnetic moment, quadrupole moment, B(E2) and B(M1) are also calculated in this work. The calculated electromagnetic observable values are also compared with the experimental data wherever it is available. In the case of $^{139}$La, the magnetic moment value calculated from shell-model for $7/2_1^{+}$ state is 1.760 $\mu_N$ whereas it is 2.7830455(9) $\mu_N$ experimentally. Likewise, in the case of $^{140}$La for $3^{-}_1$ state, this value comes out to be 0.130 $\mu_N$ from shell-model and 0.730(15) $\mu_N$ experimentally. In the case of $^{141}$La for $7/2_1^+$, the experimental value of the magnetic moment is not yet observed although the calculated value from the shell-model is 1.726 $\mu_N$. The quadrupole moment is also calculated in this work and compared with the experimental data. We are getting good shell-model results for quadrupole moment for all of the isotopes, and both shell-model and experimental values are positive which means these nuclei show prolate shape. For instance, the quadrupole moment value for $^{139}$La($7/2_1^+$) is +0.127 $eb$ theoretically, and it is +0.200(6) $eb$ experimentally. Similarly, in the case of $^{140}$La isotope for $3^{-}_1$, it is +0.152 $eb$ theoretically whereas it is +0.094(10) $eb$ experimentally. Moving forward, the experimental value of quadrupole moment for $^{141}$La($7/2_1^+$) isotope is not available, but theoretically, this value comes out to be +0.030 $eb$. In addition to these quantities, B(E2) and B(M1) are also evaluated. The shell-model results for B(M1) agree pretty well with the experimental data. For instance, B(M1) value for $^{139}$La($5/2^+_1 \rightarrow 7/2^+_1$) is 0.0003 $\mu_N^2$ theoretically whereas it is 0.00460(7) $\mu_N^2$ experimentally. Similarly, B(M1) value for $^{140}$La($2^-_1 \rightarrow 3^-_1$) is 0.813 $\mu_N^2$ theoretically which is close to the experimental value, i.e., 0.913$^{+197}_{-143}$ $\mu_N^2$. In the case of $^{141}$La($5/2^+_1 \rightarrow 7/2^+_1$), this value comes out to be 0.005 $\mu_N^2$ theoretically, and experimentally it is 0.00381$^{+34}_{-18}$ $\mu_N^2$. Further, B(E2) values are also evaluated using shell-model for all the mentioned isotopes. Corresponding results are shown in the table, the calculated result for $^{140}$La($2^-_1 \rightarrow 3^-_1$) transition is larger in comparison to the available experimental data.

After testing the wave functions through nuclear structure properties calculations, then we have calculated beta decay properties corresponding to $^{139,140,141}$Ba transitions. At first, the shell-model $\log ft$ and average shape factor values are calculated using the bare values of weak coupling constants. Similarly, the experimental average shape factor values are also extracted from the experimental $\log ft$ values, and then both shell-model and experimental results for average shape factor values are compared. This comparison is shown in Fig. \ref{shape_plot}(a). In this figure, the data points are scattered. Following this, the beta decay properties such as  $\log ft$ and average shape factor values are computed using the effective value of axial-vector coupling constant $g_A^{eff}=0.66\pm 0.03$. The comparison of these average shape factor values with the experimental average shape factor values is shown in Fig. \ref{shape_plot}(b) where the data points are less scattered.

Tables \ref{table1}, \ref{table2}, \ref{table3} show the shell-model results for beta decay properties such as the $\log ft$ values, average shape factor values reported using the effective value of axial-vector coupling constant corresponding to $^{139,140,141}$Ba isotopes, respectively. The experimental data are also given in the tables. The decay mode and spin parity of the final state along with the experimental excitation energy are also listed in these tables. In the tables, FU denotes the forbidden unique beta decay, and FNU denotes the forbidden non-unique beta decay.

In Table \ref{table1}, the shell-model $\log ft$ and the average shape factor values corresponding to beta decay of $^{139}$Ba using effective values of axial-vector coupling constant are reported and compared with the experimental ones. Since the accurate computation of $\log ft$ values requires a precise Q-value, the experimental Q-value is taken, i.e., \\
2314.6(23) keV. We are getting a good agreement between the shell-model results and experimental data for these transitions. For instance, at ground state energy, the shell-model $\log ft$ value is 6.774, which is close to the experimental value, i.e., 6.845(3). Also, at energy level 1219.047(10) keV, the shell-model $\log ft$ value is  9.332, which is close to the experimental $\log ft$ value, i.e. 9.68(1).  There are some energy levels where tentative spin parities are given experimentally. The shell-model $\log ft$ values for these states are also calculated and compared with the experimental $\log ft$ values. Based on the calculated $\log ft$ values from shell-model it might be possible to assign one particular spin parity state. For instance, at energy level 1420.491(8) keV, there are two tentative spin parities ($5/2^+,7/2^+$) given experimentally and the experimental $\log ft$ value is 7.64(1). The shell-model $\log ft$ values for both of the spin parity states are computed which comes out to be 8.708 for $5/2_3^+$ and 7.805 for $7/2_2^+$. The shell-model $\log ft$ value for $7/2_2^+$ state is closer to the experimental $\log ft$ value as compared to the other one. Thus, the $7/2_2^+$ state can be proposed at  1420.491(8) keV energy level. Further, at energy level 1578.156(14) keV, the $5/2_5^+$ state can be proposed since its shell-model $\log ft$ value is closer to the experimental value as compared to the other. Further, in other cases, the shell-model $\log ft$ values for tentative spin parity states as shown in the table, differ slightly from each other corresponding to two tentative experimental states. Thus, it is difficult to proposed one particular spin parity state at these levels from shell model.  For instance, at energy level 1558.72(3) keV, there are two tentative spin parities ($3/2^+,5/2^+$) given experimentally and the experimental $\log ft$ value is 10.21(2). The shell-model $\log ft$ values for $3/2_1^+$ state come out to be 9.371, and it is 9.292 for $5/2_4^+$. Here, both values are very close to each other. Thus, we are unable to propose one particular spin parity state at this energy level from the shell-model. A similar situation occurs for other transitions also.
	
In Table \ref{table2}, the experimental and shell-model $\log ft$ and average shape factor values using effective values of axial-vector coupling constant corresponding to the beta decay of $^{140}$Ba using experimental Q-value, i.e., 1047(8) keV are given. The shell-model $\log ft$ values match quite well with the experimental data. For instance, at energy level 29.9677(8) keV, the shell-model $\log ft$ value is 9.272$^{1u}$, which is close to the experimental $\log ft$ value, i.e., 8.91$^{1u}$(24). Similarly, at energy level 162.6585(8) keV, the shell-model $\log ft$ value is 9.166$^{1u}$, and the experimental $\log ft$ value is 9.30$^{1u}$(3). In conclusion the shell model prediction for $\log ft$ values corresponding to $^{140}$Ba $\rightarrow$ $^{140}$La transitions are fairly well.

\begin{table}
	 \centering
	\caption{Comparison between theoretical and experimental half-lives \cite{rufino2022,joshi2016,nica2018} for $^{139,140,141}$Ba isotopes using effective value of axial-vector coupling constant i.e. $(g_A^{eff}=0.66\pm 0.03)$ and $\epsilon_{MEC}=2.0$.\label{thl}} 
	\begin{tabular}{ccccccc}
 \hline\hline
		Isotope& \multicolumn{2}{c}{~~~~~Half-life}\T\B \\
		\cline{2-3}

		&  Expt. & SM   \T\B\\\hline

		$^{139}$Ba&82.93(9) min & 28.1 min \\ \T\B
		$^{140}$Ba&12.751(4) d & 1.10 d \\ \T\B
		$^{141}$Ba&18.27(7) min & 4.16  min\\\hline\hline
		
	\end{tabular}
\end{table}

  \begin{figure}
  \centering
		\includegraphics[scale=0.50]{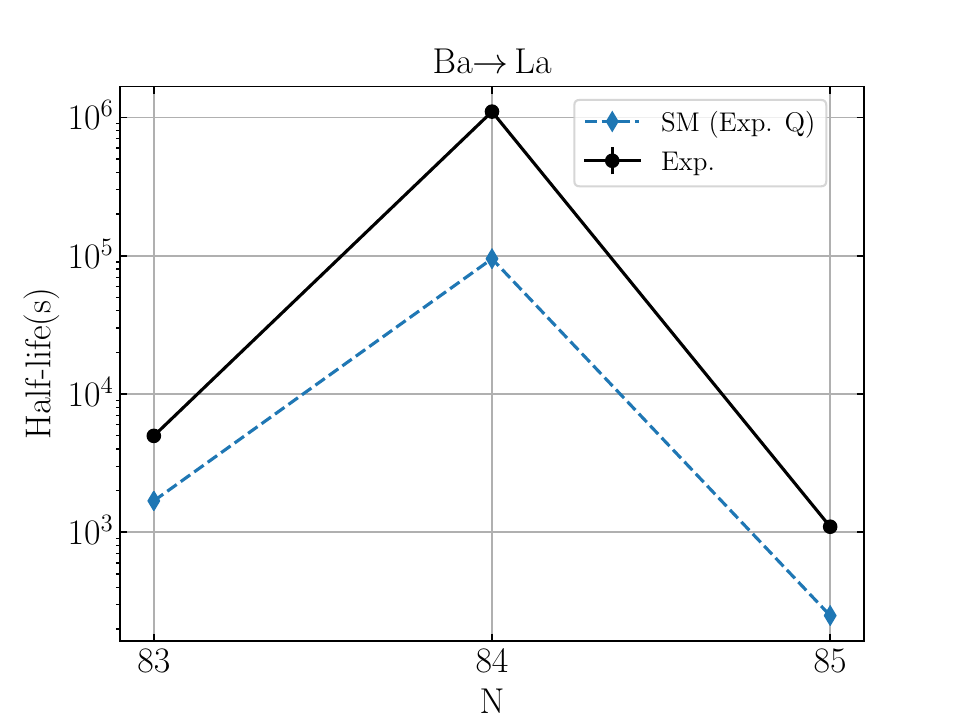}
			\caption{Comparison of experimental \cite{rufino2022,joshi2016,nica2018} and shell-model half-life
calculated using effective values of weak coupling constants.\label{hl_plot}}
\end{figure}

\begin{figure}
\centering
		\includegraphics[scale=0.50]{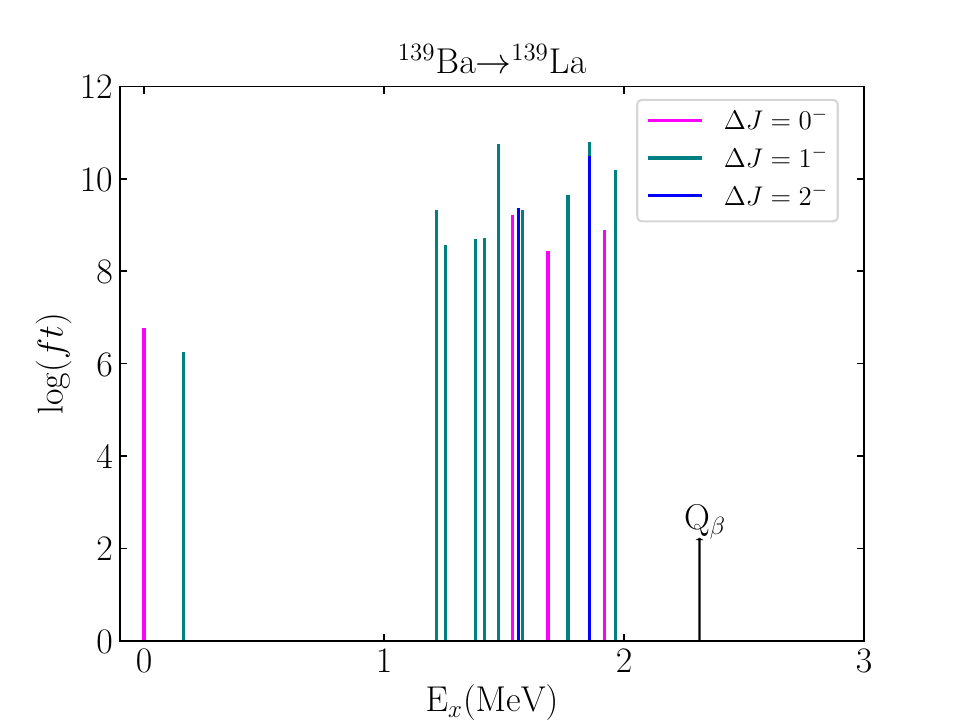}
  \includegraphics[scale=0.50]{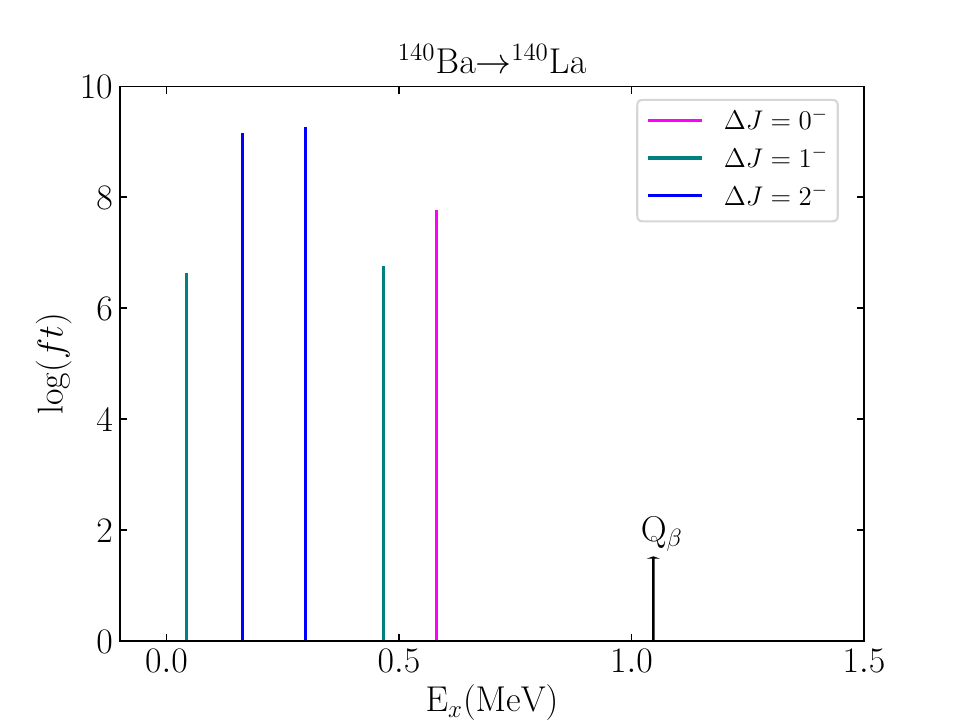}
  \includegraphics[scale=0.50]{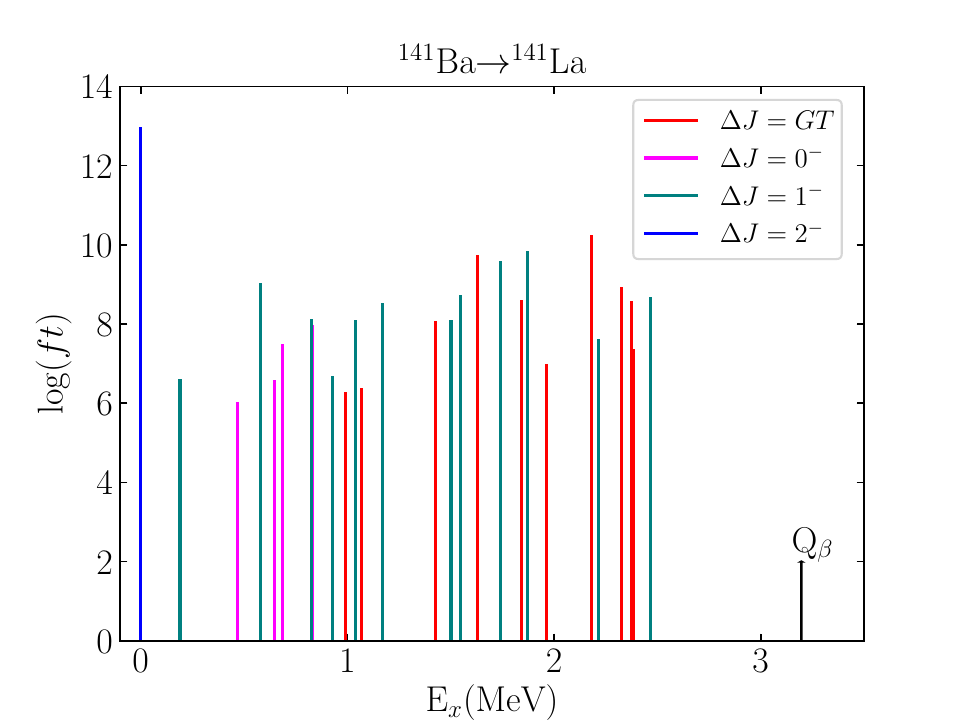}
			\caption{Shell-model results for strength functions calculated via effective values of weak coupling constants.\label{ft_plot}}
\end{figure}

\begin{table}
 \centering
\caption{Error in shell-model half-life values in comparison to experimental half-lives 
\cite{rufino2022,joshi2016,nica2018}.}\label{error1}\T\B
\begin{tabular}{cccccc}	
\hline\hline
 $\bar{r}$ &-0.726 \\ \T\B
 $\sigma$ &0.250\\ \T\B
  $10^{\bar{r}}$&0.188\\ \T\B
 $10^\sigma$& 1.776\\\hline\hline	
\end{tabular}
\end{table}

In Table \ref{table3}, the shell-model results for $\log ft$  and average shape factor values have been listed corresponding to the beta decay of $^{141}$Ba along with the experimental data from ANL \cite{rufino2022}. The experimental Q-value is used here, i.e., 3197(7) keV. The shell-model results for the $\log ft$ values are in reasonable agreement with the experimental data. For instance, the shell-model $\log ft$ value at energy level 826.42(10) is 8.121, whereas the corresponding experimental value is 8.17(3). Similarly, at energy level 831.66(8) keV, the shell-model $\log ft$ value is 7.969, which is close to the experimental $\log ft$ value, i.e., 7.720(24). There are some additional energy levels above 1171.99(9) keV experimental excitation energy for which the spin parity is not mentioned experimentally, but these states are involved in the beta decay process and thus contribute to the half-life of the nucleus. Since we can not calculate $\log ft$ values of states without knowing their spin parity with shell-model, thus, these states are excluded from the table. Because of this fact the shell-model results above 1628.16(10) keV energy levels do not agree well with the experimental data.  Hence, additional experimental data is needed.

Table \ref{thl} shows the calculated half-life values for the beta decay of $^{139,140,141}$Ba using shell-model and corresponding experimental data. The same has been plotted in Fig. \ref{hl_plot} in seconds against neutron number on the x-axis, and one can interpret from this graph that our shell-model results for half-life match fairly well with the corresponding experimental half-life values. For instance, the shell-model half-life value for $^{141}$Ba decay is 4.16 min which is not very far from the corresponding experimental half-life value, i.e., 18.27(7) min.

Fig. \ref{ft_plot} shows strength function plots \cite{borzov2011} of $^{139,140,141}$Ba with respect to the excitation energy of daughter nuclei. These plots also display the corresponding Q-value. Moreover, the contribution of allowed and forbidden transitions can be studied through these plots. In $^{141}$Ba decay, there are some GT transitions involved, i.e., $^{141}$Ba($3/2^-$) $\rightarrow$ $^{141}$La($3/2^-$). These GT transitions include the transition of the neutron from $0h_{9/2}$ orbital to the $0h_{11/2}$ proton orbital and these transitions are typically seen towards higher excitation energy.

We have also calculated the total deviation \cite{yoshida2018,moller2003,marketin2016} in the shell-model half-life values from the experimental half-lives. Suppose $r$ is the measure of deviation, which can be calculated from the theoretical and experimental total half-lives by the expression

\begin{equation}
	r=\log_{10}(T_{1/2}^{\rm{SM}}/T_{1/2}^{\rm{expt}}),
\end{equation}
where $T_{1/2}^{\rm{SM}}$ is the total half-life value calculated from shell-model and the experimental half-life value is denoted by $T_{1/2}^{\rm{expt}}$. The total half-life can be obtained from partial half-lives by using the expression,
\begin{equation}
\frac{1}{T_{1/2}}=\sum_k \frac{1}{t_{1/2}^{(k)}},
\end{equation}
where summation is given over the total number of transitions ($k$) involved in the beta-decay.
The deviation is calculated in the logarithmic scale, and $\bar{r}$ and $\sigma$ are termed as the mean value in the deviation and its standard deviation, respectively which are given in the form

\begin{equation}
	\bar{r}=\frac{1}{n}\sum_{i=1}^n r_i,
\end{equation}

\begin{equation}
	\sigma=\Big[\frac{1}{n}\sum_{i=1}^n(r_i-\bar{r})^2\Big]^{1/2},
\end{equation}

where $n$ is the total number of beta-decaying isotopes considered in this work for which the deviation is being calculated. The mean deviation and standard deviation should be close to zero as these are calculated in the logarithmic scale. These quantities are also given in the power of 10 in the table. From Table \ref{error1}, it can be concluded that our shell-model results are pretty well according to the experimental data as $\sigma=0.250$ which means these values approach zero. \\

\section{Conclusion}\label{conclusion}
Motivated by recent experimental data for beta decay properties of $^{141}$La from Argonne National Laboratory (ANL),
in this work, the nuclear structure and the beta decay properties are calculated using shell-model and compared with the corresponding experimental data. The nuclear properties such as energy spectra and electromagnetic observables are computed using jj56pnb interaction. The energy spectra of ground and excited states of $^{139,140,141}$La are also shown. 
The shell-model calculations correctly reproduce the spin parity of the ground state of all the mentioned isotopes and also reasonably describe the overall energy spectra. In addition, the electromagnetic observables such as magnetic moment, quadrupole moment, B(M1) and B(E2) are also calculated 
After calculating nuclear structure properties, the beta decay properties are calculated using these wave functions. At first, the $\log ft$ and average shape factor values are computed using bare values of coupling constants.  Further, the effective values of axial-vector coupling constant is used to calculate the $\log ft$ values and average shape factor values which are listed and compared with the experimental data. 
Now, the total half-life values are calculated using these $\log ft$ values. In order to establish a better comparison, the error in the calculated half-life values is computed with respect to the experimental half-life values.  Overall our results are fairly close to the experimental data.
\section*{Acknowledgement}

This work is supported by a research grant from SERB (India), CRG/2022/005167.   
S. S. would like to thank CSIR-HRDG (India) for the financial support for her Ph.D. thesis work. We acknowledge the National Supercomputing Mission (NSM) for providing computing resources of ‘PARAM Ganga’ at the Indian Institute of Technology Roorkee.

\end{document}